\documentclass[conference]{IEEEtran}
\usepackage{cite}
\usepackage{amsmath,amssymb,amsfonts}
\usepackage{algorithmic}
\usepackage{graphicx}
\usepackage{textcomp}
\usepackage{xcolor}
\usepackage{booktabs}
\usepackage{siunitx}
\usepackage{subcaption}
 \usepackage{booktabs}

\usepackage{amsthm}
\usepackage{hyperref}
\usepackage{float}
\usepackage{tikz}
\usetikzlibrary{shapes,arrows,positioning,calc,fit}

\newtheorem{theorem}{Theorem} [section]

\theoremstyle{definition}

\newtheorem{example}[theorem]{Example}

\begingroup\expandafter\expandafter\expandafter\endgroup
\expandafter\ifx\csname pdfsuppresswarningpagegroup\endcsname\relax
\else
\fi

\def\BibTeX{{\rm B\kern-.05em{\sc i\kern-.025em b}\kern-.08em
    T\kern-.1667em\lower.7ex\hbox{E}\kern-.125emX}}

\begin{document}

\title{Quality of service based radar resource management for interference mitigation
}

\author{\IEEEauthorblockN{Sebastian Durst}
	\IEEEauthorblockA{\textit{Fraunhofer-Institut f\"ur Hochfrequenz-} \\
		\textit{physik und Radartechnik FHR}\\
		Wachtberg, Germany \\
		sebastian.durst@fhr.fraunhofer.de
	}
	\and
	\IEEEauthorblockN{Pascal Marquardt}
	\IEEEauthorblockA{\textit{Fraunhofer-Institut f\"ur Hochfrequenz-} \\
		\textit{physik und Radartechnik FHR}\\
		Wachtberg, Germany \\
		pascal.marquardt@fhr.fraunhofer.de}
	\and
	\IEEEauthorblockN{Stefan Br\"uggenwirth}
	\IEEEauthorblockA{\textit{Fraunhofer-Institut f\"ur Hochfrequenz-} \\
		\textit{physik und Radartechnik FHR}\\
		Wachtberg, Germany \\
		stefan.brueggenwirth@fhr.fraunhofer.de}
}

\maketitle

\begin{abstract}
	An intelligent radar resource management is an essential building block of any modern radar system.
	The quality of service based resource allocation model (Q-RAM) provides a framework for profound and quantifiable decision making but lacks the flexibility necessary for optimal mitigation strategies in the presence of interference.
	We define an extension of the Q-RAM based radar resource management framework with an intelligent interference handling capability using various mitigation methods.
	The approach incorporates virtual time resources and alternative task configurations to compute near-optimal solutions in the presence of interference.
	The provided experimental results demonstrate a significant improvement over traditional strategies.
\end{abstract}


%%%%%%%%%%%%%%%%%%%%%%%%%%%%%%%%%%%%%%%%

% Tikz style
\tikzset{
	block/.style = {draw, fill=white, rectangle, minimum height=3em, minimum width=3em},
	tmp/.style  = {coordinate}, 
	sum/.style= {draw, fill=white, circle, node distance=1cm},
	input/.style = {coordinate},
	output/.style= {coordinate},
	pinstyle/.style = {pin edge={to-,thin,black}
	}
}	

\tikzstyle{decision} = [diamond, draw, %fill=blue!20, 
text width=6em, text badly centered, node distance=3cm, inner sep=0pt, font=\large]
\tikzstyle{block} = [rectangle, draw, %fill=blue!20, 
text width=10em, text centered, rounded corners, minimum height=4em, font=\large]
\tikzstyle{line} = [draw, -latex']
\tikzstyle{cloud} = [draw, ellipse, %fill=red!20,
node distance=3cm, minimum height=2em]

%%%%%%%%%%%%%%%%%%%%%%%%%%%%%%%%%%%%%%%%%

\section{Introduction}
\label{sec:intro}

Resource management (the prioritisation, resource allocation and scheduling of tasks) is an essential part of modern radar systems
since they can simultaneously perform potentially conflicting functions
and even more so of any cognitive setup.
Tasks can usually be executed in a multitude of configurations with different resource requirements and resulting utility. One goal of the resource management
is to select parameters such that the overall system performance is optimised while respecting resource constraints.
A mathematical framework describing this problem is known as \emph{Quality of service based resource allocation model (Q-RAM)}
\cite{Raj1997, Ghosh2006}.

A huge problem in radar is the existence of other emitters in the field of view of a radar like jammers, other radars or broadcasting stations. These can cause electromagnetic interference and thus negatively impact the quality of task execution of the system.
There are several potential strategies to mitigate these effects like
the choice of execution time,
the choice of frequency,
the use of adaptive beamforming, or
the change of search pattern and direction.
These techniques have to be considered and chosen in the resource management module of the radar system to assess their global impact on system utility.
If the interference is permanent or at least potentially constant,
i.e.\ caused by an interferer that does not necessarily have to be constant but has to be expected to interfere at any given time,
it is straight-forward to modify the Q-RAM framework to handle the situation.
The underlying performance models have to be able to take into account the impact of the interference but there are no further consequences for resource management or other decision making processes.
A non-permanent interferer with known or learnt characteristics, e.g.\ a broadcasting station, allows for an adaptive decision making where mitigation strategies are chosen depending on the time of execution.
In classical Q-RAM, resources are allocated to tasks without knowing their exact execution time. Hence, the framework has to be extended to ensure performance in the presence of non-constant interference.

The paper is organised as follows.
In Section~\ref{sec:qram_problem} the Q-RAM problem is formulated and the classical solution approach is briefly introduced in Section~\ref{sec:classical_solution}.
Our modified Q-RAM based resource management architecture for interference mitigation is presented in
Section~\ref{sec:architecture}.
We then discuss a specific optimisation algorithm in 
Section~\ref{sec:algo}.
We provide experimental results demonstrating the effectiveness of the algorithm in 
Section~\ref{sec:sim} before concluding the paper in
Section~\ref{sec:conclusion}.

\section{The Q-RAM problem}
\label{sec:qram_problem}

We will briefly introduce the Q-RAM problem in this section.% (cf.~\cite{qram_rl}).
The goal of the resource allocation module in a multifunction radar is to maximise the 
\emph{utility} of a set of radar tasks by selecting \emph{operational parameters} (e.g.\ waveform, dwell period or the choice of tracking filter)
while adhering to certain \emph{resource constraints} (e.g.\ radar bandwidth, power) and taking into account the \emph{environmental conditions}, i.e.\ situational data.
A specific choice of operational parameters together with the environmental conditions determine the (expected) \emph{quality} of a task, which is usually task-type related and allows for easier, interpretable user control. Quality and situational data then define the task utility.
For $\{\tau_1,\ldots,\tau_n\}$ a set of radar tasks, $k$ types of resources with resource bounds $R_1,\ldots,R_k$ and environmental conditions $e$, the problem can be formulated as
\begin{align}
	\max_{\phi = (\phi_1,\ldots,\phi_n)}& u(\phi, e)\\
	\textrm{s.t. } \forall j=1,\ldots,k\;\;& \sum_{i=1}^n \big(g_i(\phi_i)\big)_j \leq R_j,
\end{align}
where $\phi_i$ is a configuration for task $\tau_i$, $u$ the system utility and $g_i$ functions mapping task configurations to their resource requirements
(see \cite{qram_rl} for a more detailed description).

\begin{figure*}[tb] 
	\centering
	\includegraphics[width=0.95\textwidth]{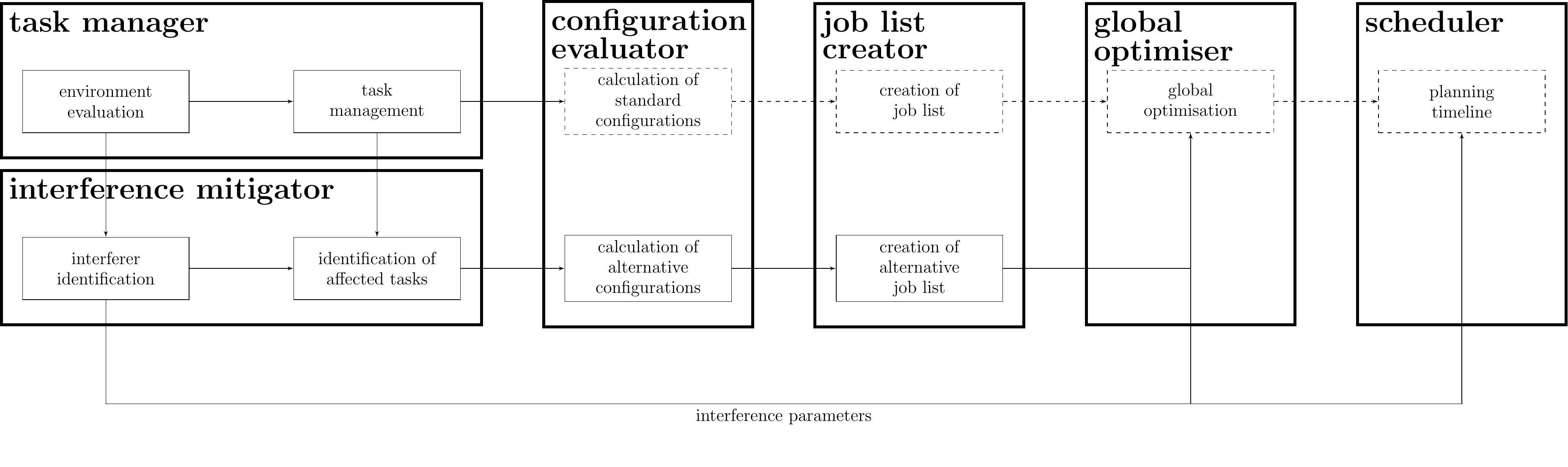}
	\caption{Block diagram of the proposed resource management architecture. The classical Q-RAM framework is indicated by dashed lines.} \label{fig:block_diagram}
\end{figure*}

\section{The classical approach}
\label{sec:classical_solution}

In this section we will briefly describe the approximative solution strategy to the Q-RAM problem proposed
in \cite{Raj1997, Ghosh2006}.
A notable alternative approach is given in \cite{CharlishPhD}.
First, all possible task configurations are generated and evaluated on a per task basis, i.e.\ their resulting utility and resource requirement are computed. In case of multiple resources, a scalar proxy, called \emph{compound resource}, is used.
Then a subset of configurations with a high utility for various resource levels is pre-selected and stored
in a list ordered by increasing compound resource requirement, which we will refer to as \emph{job list}.
The job lists of all tasks are given to a global optimiser, which iteratively allocates resources to the task offering the best utility-to-resource-ratio provided sufficient resources are available.
If a task is selected and assigned additional resources, we say that the task is \emph{upgraded}.
After resource allocation, the selected jobs (i.e.\ tasks with a chosen configuration) are scheduled by a scheduler.

\section{The Adapted Architecture}
\label{sec:architecture}

A standard Q-RAM based resource management as described in the previous section is not able to handle the effects of interference adequately, as the scheduling is subordinated and the resource allocation algorithm lacks information on particular task execution times.
We propose an extension of the classical Q-RAM framework for interference mitigation as described by the block diagram in Figure~\ref{fig:block_diagram}.
The module \emph{interference mitigator}  receives information on the environment and a list of tasks that are to be considered for the next planning interval. The module then identifies interferers and their characteristics and checks which tasks are potentially affected. This information is then used to create alternative configurations and job lists for these tasks in the \emph{configuration evaluator} and the \emph{job list creator} to represent the use of mitigation techniques.
Furthermore, the interference parameters are sent to the \emph{global optimiser} and the \emph{scheduler}, which can  divide the radar timeline into interference-free and interference-present intervals that are represented as virtual resources during resource allocation.
This allows for a flexible and cognitive decision making on a per task basis for the use of interference mitigation or interference-free scheduling.

It is worth noting that the adapted architecture can be easily implemented in existing Q-RAM based systems.

\section{The Optimisation Algorithm}
\label{sec:algo}

Assume a single cooperative or otherwise known (e.g.\ learnt) interferer that broadcasts according to a known pattern interfering with radar tasks. The interference does not have to be periodic but can occur at any known points in time.
Tasks can be divided into two classes: tasks that are prone to interference and tasks that are interference-free because of their beam direction or used frequency.
In the standard Q-RAM algorithm, the input to the global optimiser in the resource allocation step is a job list for every task.
In our modification, tasks prone to interference need to have two separate job lists, a \emph{standard} and an \emph{alternative} job list.
The standard job list consists of jobs in their standard configuration, i.e.\ the usual configurations that would be interfered. The alternative job list contains alternative configurations adapted to make use of suitable interference mitigation strategies like a frequency change or adaptive beamforming.
In case no mitigation strategies are available, the alternative configurations could also be the standard configurations with a decreased utility due to interference.

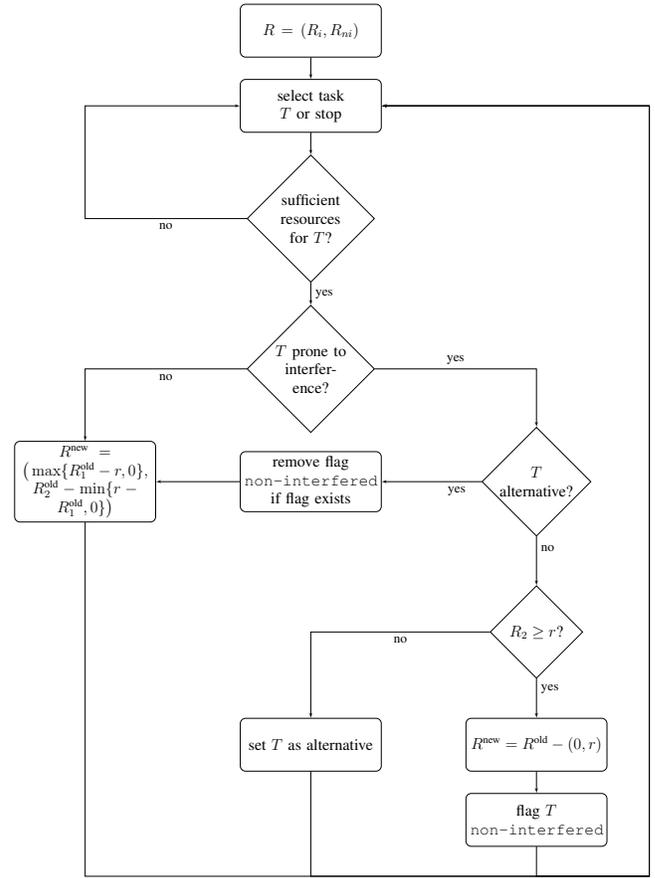
\begin{figure}[htb] 
	\centering
	\scalebox{0.5}{
	\begin{tikzpicture}[node distance = 2cm, auto]
		% Place nodes
		\node [block] (init) {$R=(R_i, R_{ni})$};
		\node [block, below of=init] (select)  {select task $T$ or stop};
		\node [decision, below of=select] (res)  {sufficient resources for $T$?};
		\node [decision, below of=res, node distance=4cm] (interfer) {$T$ prone to interference?};
		\node [block, below of=interfer, node distance=3cm] (remflag) {remove flag \texttt{non-interfered} if flag exists};
		\node [block, left of=remflag, node distance=6cm] (updatenormal)  {$R^{\textrm{new}} = $\\$ \big( \max\{R^{\textrm{old}}_1 - r, 0\} ,$\\$ R^{\textrm{old}}_2 - \min\{r - R^{\textrm{old}}_1, 0\} \big)$};
		\node [decision, right of=remflag, node distance=6cm] (alternative) {$T$ alternative?};
		\node [decision, below of=alternative, node distance=4cm] (r2) {$R_2\geq r$?};
		\node [block, below of=r2, node distance=3cm] (updateinterfer) {$R^{\textrm{new}} = R^{\textrm{old}} - (0, r)$};
		\node [block, left of=updateinterfer, node distance=6cm] (setalt) {set $T$ as alternative};
		\node [block, below of=updateinterfer] (flag)  {flag $T$ \texttt{non-interfered}};
		\node[coordinate,below of=setalt, node distance=3.5cm] (invisiblebottom) {};
		\node[coordinate,right of=invisiblebottom, node distance=9cm] (invisibleright) {};
		\node[coordinate, left of=select, node distance=6cm] (invisibleleft) {};
		
		% Draw edges
		\path [line] (init) -- (select);
		\path [line] (select) -- (res);
		\path [line] (res) -- node {yes}(interfer);
		\path [line] (res.west) -| node [near start] {no} (invisibleleft) -- (select.west);
		\path [line] (interfer) -| node [near start] {no}(updatenormal);
		\path [line] (interfer) -| node [near start] {yes}(alternative);
		\path [line] (alternative) -- node [near start] {yes}(remflag);
		\path [line] (remflag) -- (updatenormal);
		\path [line] (alternative) -- node [near start] {no}(r2);
		\path [line] (r2) -| node [near start] {no}(setalt);
		\path [line] (r2) -- node [near start] {yes}(updateinterfer);
		\path [line] (updateinterfer) -- (flag);
		\path [line] (updatenormal) |- (invisiblebottom) -- (invisibleright) |- (select);
		\path [line] (setalt) -- (invisiblebottom) -- (invisibleright) |- (select);
		\path [line] (flag) |- (invisibleright) |- (select);
	\end{tikzpicture}
	}
	\caption{Flow chart of the global optimisation step.} \label{fig:flow_chart}
\end{figure}

As the interference pattern is known, we can determine the time share with interference present and divide the radar time resource accordingly into two components $R_i$ (interference possible) and $R_{ni}$ (no interference). These will be treated (almost) like separate resources in the resource allocation process.
A flow chart of the modified algorithm is given in Figure~\ref{fig:flow_chart}. The main idea is that tasks that are not prone to interference use resource $R_i$ if possible while potentially interfered tasks use $R_{ni}$. In case a task prone to interference is selected for upgrade but $R_{ni}$ is insufficient, the task is set to its alternative job list and can then be chosen in an interference-mitigated configuration in a subsequent iteration.
In more detail, the modified algorithm operates as follows:
\begin{itemize}
	\item All tasks are set to the base configuration of their standard (non-interfered) job list.
	\item As in standard Q-RAM, the task with the highest resulting utility to resource ratio is selected for upgrade.
	\begin{itemize}
		\item If this task is not prone to interference, it is upgraded and its time resource is subtracted from the $R_{i}$ budget.
		\begin{itemize}
			\item In case the $R_{i}$ budget is non-sufficient, the $R_{ni}$ budget can be used to accommodate the surplus of resource need.
		\end{itemize}
		\item If the task is prone to interference, it is checked whether the $R_{ni}$ budget is sufficient for the upgrade.
		\begin{itemize}
			\item If so, the task is upgraded as usual using $R_{ni}$ and it is flagged \texttt{non-interfered}.
			\item If not, the task is changed to its alternative job list and a new task is selected for upgrade.
		\end{itemize}
	\end{itemize}
	\item The above step is repeated until no further upgrades are possible.
\end{itemize}

The described modification has the effect that the scheduler can choose the execution time of tasks flagged \texttt{non-interfered} such that no interference is present. Hence, high profile tasks can be executed in their best possible configuration and do not have to fall back to potentially utility-decreasing mitigation strategies only because interference is partly present in the planning interval.

\begin{example}
	\label{ex:algo}
	Consider a resource management problem with three tasks $T_1, T_2$ and $T_3$ and a single jammer that potentially interferes with $T_1$ and $T_2$ and is transmitting 60 percent of the time. Time is the only resource under consideration. Assume that all tasks have a non-execution base configuration with no resource requirements and zero utility.
	The tasks all have a standard configuration denoted by $c_1$, $c_2$ and $c_3$, respectively, while the interference-prone tasks $T_1$ and $T_2$ each have an additional alternative configuration $c_{1a}$ and $c_{2a}$, respectively. The alternative configurations make use of a frequency change and are not interfered by the jammer. The resource requirements and utilities of the configurations are listed in Table~\ref{tab:configs}.
	As the jammer is active 60 percent of the time, we set $R_i = 0.6$ and $R_{ni} = 0.4$.
	The global optimiser first selects $T_2$ and upgrades it from its base configuration to $c_2$, as it has the highest ratio of utility to resource. As the task can be interfered in this standard configuration, the interference-free resource $R_{ni}$ is used.
	Next, task $T_3$ is upgraded and resource $R_i$ is used. After these two upgrade steps, the resource budget amounts to $(R_i, R_{ni}) = (0.3, 0.1)$. An upgrade of task $T_1$ to its standard configuration is not possible, as there are insufficient interference-free resources. However, the task can be performed in its alternative configuration $c_{1a}$.
	Thus, the resource allocation process yields a total utility of $2.1$,
	whereas a traditional system always relying on alternative configurations in the presence of an interferer would only amount to a total utility of $1.8$ in this example.
	
	\begin{table}[t]
		\centering	
		\caption{Task configurations for Example~\ref{ex:algo}.}
		\label{tab:configs}
		\begin{tabular}[t]{lrr}
			\toprule
			&$r$&$u$\\
			\midrule
			$c_1$&$0.3$&$0.6$\\
			$c_{1a}$&$0.3$&$0.4$\\
			$c_2$&$0.3$&$0.9$\\
			$c_{2a}$&$0.3$&$0.6$\\
			$c_3$&$0.3$&$0.8$\\
			\bottomrule
		\end{tabular}
	\end{table}%
\end{example}
 
Choosing configurations in a greedy way during global optimisation can lead to suboptimal results. For example, this can be mitigated by considering multiple parallel choices in a tree-like structure, as will be presented in a subsequent publication.
 
Note that the algorithm can be easily adapted to situations with multiple sources or areas of interference by defining multiple (possibly overlapping) virtual time resources.

\begin{figure}[htb] 
	\centering
	\includegraphics[width=.35\textwidth]{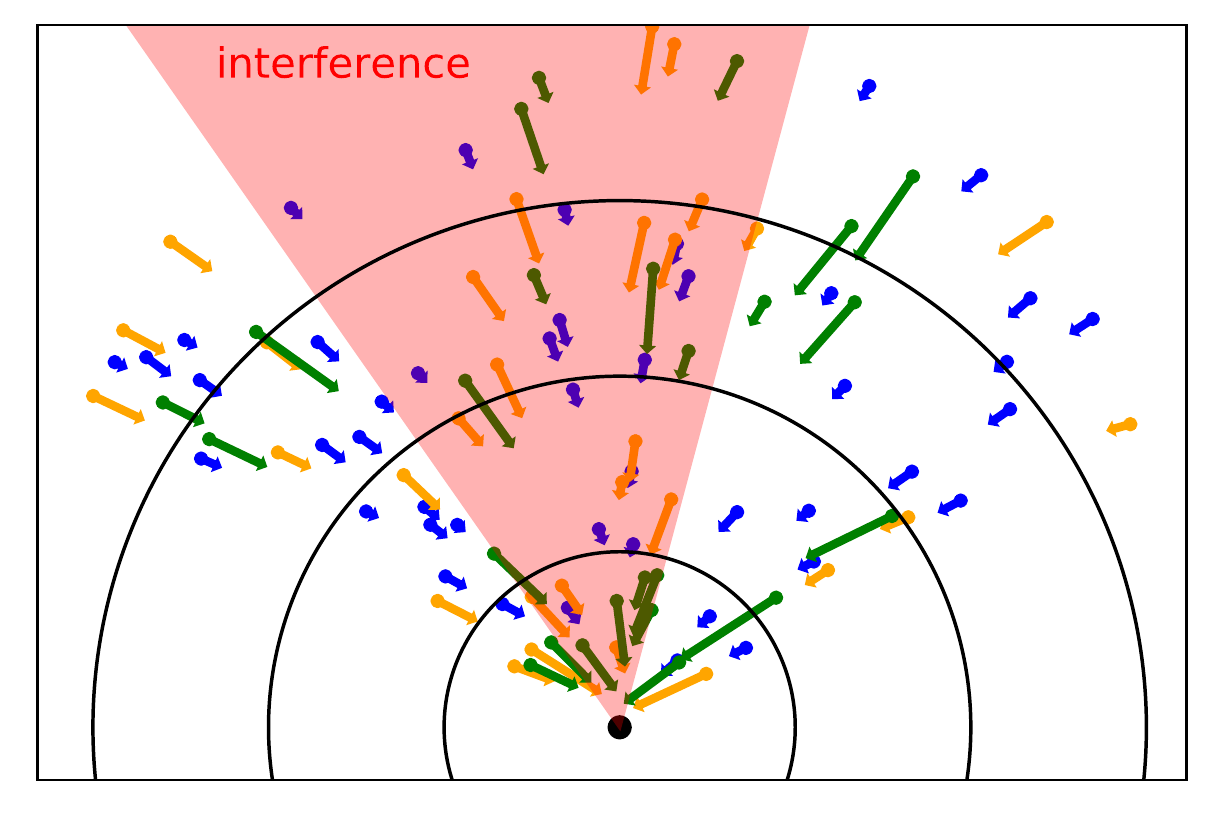}
	\caption{Scenario overview. Target position and velocity with the colour encoding the target type. Interference is present in the red region.} \label{fig:scenario}
\end{figure}

\section{Simulation Results}
\label{sec:sim}

\begin{figure}[htb] 
	\centering
	\includegraphics[width=.5\textwidth]{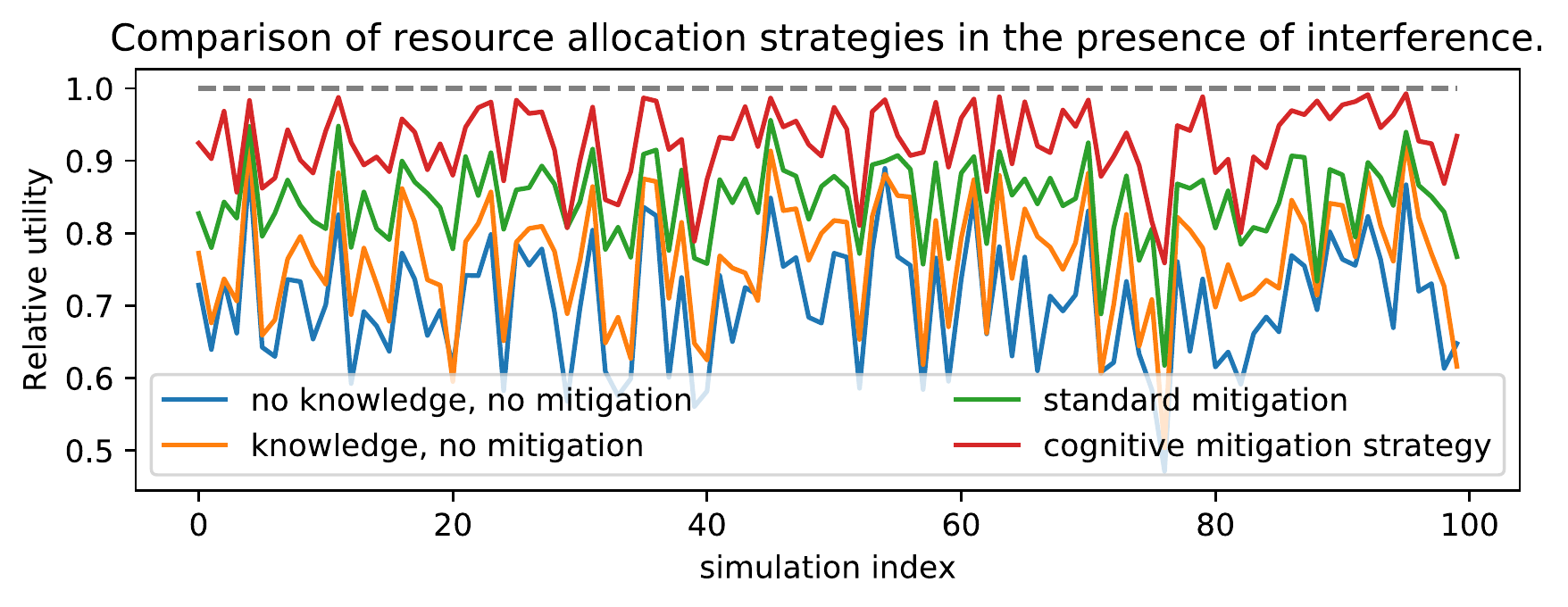}
	\caption{Simulation results of single runs.} \label{fig:comp}
\end{figure}

\begin{figure}[h!] 
	\centering
	\includegraphics[width=.5\textwidth]{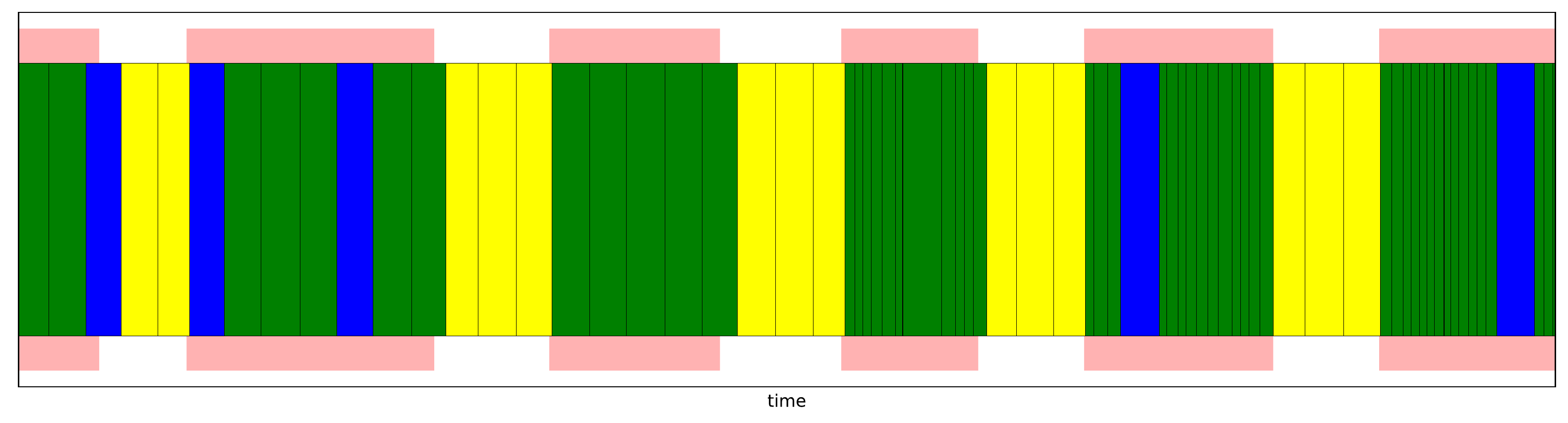}
	\caption{Excerpt of a resulting schedule. Red areas represent times when the interferer is active. The coloured blocks represent scheduled tasks. Green tasks are not prone to interference, yellow tasks are in a standard configuration and blue tasks are in an alternative configuration.} \label{fig:schedule}
\end{figure}

We demonstrate the effectiveness of the proposed framework in the following tracking scenario based on \cite{Ghosh2006} and extended to the interference problem.
In particular, the utility and resource functions correspond to the ones presented there
(where the quality is inversely related to the tracking error).
A single instance of the scenario is restricted to a single planning cycle in the Q-RAM process.
100 targets of three target types of different dynamics are distributed over 90 degrees in azimuth in the radar's field of view.
A single interferer can potentially interfere with radar signals in an area of 50 degrees in azimuth (see Figure~\ref{fig:scenario}).
The interference pattern is chosen randomly in every run of the simulation such that it is active for 70\% of the time, but assumed to be fully known to the radar.
Tracking tasks in a standard configuration result in 0\% to 30\% of their regular utility when interfered, alternative configurations generate 30\% to 90\% of the regular utility. These percentages are chosen randomly for each task individually. Hence, different strength of interference and mitigation methods are simulated implicitly.
We compare the following general strategies during resource allocation (see Figure~\ref{fig:comp}):
\begin{itemize}
	\item No knowledge of the interference, no mitigation applied.
	\item Knowledge of the interference without mitigation. The system prefers tasks that are not prone to interference.
	\item Choice of mitigation method for all tasks that are prone to interference.
	\item Proposed cognitive strategy that can make use of mitigation and interference-free scheduling.
\end{itemize}
The results were computed in a Monte Carlo experiment with 100 runs. The scenario without the presence of the interferer served as a baseline for the comparison.
The results show that the proposed strategy clearly outperforms the other strategies.
It reached an average of 93\% utility when compared with the interference-free scenario and is also the most stable with a standard deviation of 0.05 
(see Table~\ref{tab:results}).
The standard strategy that chooses alternative configurations for interference mitigation for all tasks prone to interference follows with a significantly lower average utility level of 84\%.
When comparing the single runs with the worst performance, the discrepancy increases (76\% versus 62\%).
As expected, the strategies without active mitigation perform worst.
The proposed cognitive mitigation strategy did not only yield the best performance on average but it provided the best results in every single run of the simulation (see Figure~\ref{fig:comp}).
An excerpt of a resulting schedule is illustrated in Figure~\ref{fig:schedule}.

\begin{table}
	\centering
	\caption{Mean, standard deviation, minimum and maximum of the achieved normalised utility per allocation strategy.}\label{tab:results}
	\begin{tabular}{lrrrr}
		\toprule
		{} &  mean &  std.\ dev. &  min &  max \\
		\midrule
		no knowledge, no mitigation   &  0.70 &       0.08 & 0.47 & 0.89 \\
		knowledge, no mitigation      &  0.76 &       0.08 & 0.50 & 0.93 \\
		standard mitigation           &  0.84 &       0.06 & 0.62 & 0.96 \\
		cognitive mitigation strategy &  0.93 &       0.05 & 0.76 & 0.99 \\
		\bottomrule
	\end{tabular}
\end{table}


\section{Conclusion}
\label{sec:conclusion}

An extension of a Q-RAM based radar resource management architecture has been presented to enable a radar system to intelligently and adaptively choose mitigation strategies in the presence of interference.
A specific global optimisation algorithm has been discussed in detail and its effectiveness has been validated by simulations.
In the case of a single known source of interference, the proposed solution has been shown to perform close to the non-interfered case.

%%%%%%%%%%%%%%%%%%%%%%%%%%%%%%%%%%%%%%%%

\bibliographystyle{IEEEtran}
\bibliography{IEEEabrv,lit}

\end{document}